\documentclass[sn-mathphys,Numbered,pdflatex]{sn-jnl}%
\usepackage{indentfirst}
\usepackage{graphicx}%
\usepackage{multirow}%
\usepackage{amsmath,amssymb,amsfonts}%
\usepackage{amsthm}%
\usepackage{mathrsfs}%
\usepackage[title]{appendix}%
\usepackage{xcolor}%
\usepackage{textcomp}%
\usepackage{manyfoot}%
\usepackage{booktabs}%
\usepackage{amsmath}%
\usepackage[range-units=single]{siunitx}%
\usepackage{cleveref}
\usepackage{commath}

\DeclareMathOperator{\median}{median}
\DeclareMathOperator{\mad}{mad}
\DeclareMathOperator{\argmax}{argmax}
\renewcommand\vec{\mathbf}
    
\begin{document}

\title{Signatures of Thermal and Electrical Crosstalk in a Microwave Multiplexed Hard X-ray Transition Edge Sensor Array}

\author[1,2]{\fnm{Panthita} \sur{Triamkitsawat}}\email{panthitat@uchicago.edu}
\author*[1]{\fnm{Tejas} \sur{Guruswamy}}\email{tguruswamy@anl.gov}
\author[1,2]{\fnm{Orlando} \sur{Quaranta}}\email{oquaranta@anl.gov}
\author[1]{\fnm{Lisa} \sur{Gades}}\email{gades@anl.gov}
\author[1]{\fnm{Umeshkumar} \sur{Patel}}\email{upatel@anl.gov}
\author[1]{\fnm{Antonino} \sur{Miceli}}\email{amiceli@anl.gov}

\affil*[1]{\orgdiv{X-ray Science Division}, \orgname{Argonne National Laboratory}, \orgaddress{\street{S. Cass Ave}, \city{Lemont}, \postcode{60439}, \state{IL}, \country{USA}}}

\affil[2]{\orgdiv{Pritzker School of Molecular Engineering}, \orgname{The University of Chicago}, \orgaddress{\street{S. Ellis Avenue}, \city{Chicago}, \postcode{60637}, \state{IL}, \country{USA}}}

\abstract{We investigate the crosstalk between Transition-Edge Sensor (TES) pixels in a 24-pixel hard X-ray spectrometer array fabricated at the Advanced Photon Source, Argonne National Laboratory. Analysis shows thermal cross talk, possibly associated with insufficient thermalization, and rare but larger in magnitude electrical crosstalk between specific perpetrator-victim pixel combinations, potentially due to defects in the bias wiring or microwave multiplexing circuit. We use a method based on group-triggering and averaging to isolate the crosstalk response despite only having access to X-ray photon illumination uniform across the entire array. This allows us to identify thermal and electrical crosstalk between pixel pairs in repeated measurements to the level of 1 part in 1000 or better. In the array under study, the magnitude of observed crosstalk is small but comparable to the resolving power of this pixel design ($E/\Delta{}E \sim$ 1000 at \SI{20}{keV}) and so potentially responsible for a degradation in energy resolution of the array at high incident photon rates. Having proven the methods to identify and quantify crosstalk in our setup, we can now consider mitigations.}

\maketitle

\section{Introduction}\label{intro}

Superconducting Transition-Edge Sensor (TES) quantum microcalorimeters offer significantly better energy resolution compared to semiconductor detectors for X-ray photons. This improved energy resolution and near-unity quantum efficiency enable the distinguishing of small or overlapping peaks in X-ray emission spectra, rendering them an invaluable tool in the synchrotron and X-ray science communities~\cite{Doriese2017}. At the Advanced Photon Source (APS), the hard X-ray synchrotron of Argonne National Laboratory, we have built a hard X-ray TES spectrometer instrument which has been used to run high-resolution spectroscopy~\cite{Guruswamy2020,Guruswamy2021} and Compton scattering experiments~\cite{Patel2022} at the beamline 1-BM-C.

To achieve the collecting area and total count rate capability needed for synchrotron applications multiple pixels are required, organized into two-dimensional arrays. Since the wiring necessary to directly read out such arrays would result in excessive heat load for typical cryostats, a multiplexing scheme is necessary. In our instruments at the Advanced Photon Source, we use superconducting microwave-frequency multiplexing chips from the Quantum Sensors Group of the National Institute of Standards and Technology (NIST)~\cite{Mates2008}. In this scheme each TES pixel is coupled to an RF SQUID and a superconducting microwave resonator. The readout electronics we have available can continuously sample up to 128 pixels over a single coaxial line.

Many of the TES arrays characterized during the development of the APS spectrometer show an energy resolution with an unwelcome dependence on the photon energy and count rate. One possible explanation for this is the presence of crosstalk between the pixels, which can affect the baseline signal and therefore the estimation of pulse height and photon energy~\cite{Mates2019,Vaccaro2022}. Due to the tight physical spacing between each pixel and their bias wiring, the proximity in frequency of the resonators, as well as nonideal behavior of components in the RF readout chain, arrays of this type are potentially susceptible to both thermal and electrical crosstalk.

Unfortunately we do not have the hardware capability to illuminate or bias individual pixels of our arrays and systematically study their response, as performed in e.g. Ref. \cite{Wang2021}. Our pixels are biased in series by a single pair of wires, all X-ray sources currently available to us distribute photons approximately uniformly over our sensor active area, and systematically fabricating and aligning single-pixel apertures is too time-consuming for large arrays. Therefore, we decided to determine if post-acquisition analysis is sufficient to quantify crosstalk while only using our standard photon sources, similar to the methods of Refs. \cite{Miniussi2020,Taralli2021}. As a first step we started with a representative but relatively small chip -- a 24-pixel array (fabricated at ANL) compatible with our laboratory setup. In this paper, we present both the development of these methods and analysis results on this 24-pixel TES array.

\section{Methods}\label{methods}

\subsection{Device Design}

\begin{figure}
 \centering
 \includegraphics[width=125mm]{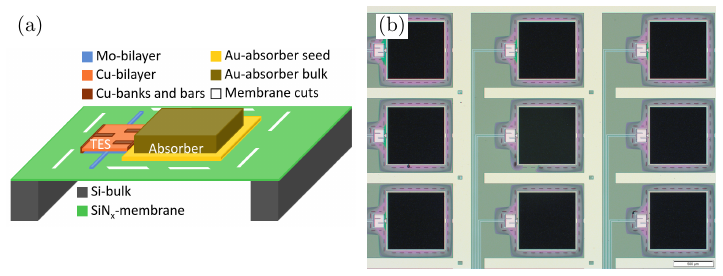}
 \caption{(a) Cross-sectional schematic showing the ``sidecar'' design with TES and absorber on a perforated SiN$_{x}$ membrane. (b) Optical microscope image showing a 3$\times$3 pixel section of a completed array of this design, fabricated at ANL.\label{fig:schematic}}
\end{figure}

The 24-pixel array under test is composed of a $6 \times 4$ grid of identical TES pixels made of \SI{170}{\um} $\times$ \SI{130}{\um} superconducting molybdenum-copper bilayers and \SI{770}{\um} $\times$ \SI{770}{\um} sidecar absorbers, all on a silicon nitride (SiN$_x$) membrane. \Cref{fig:schematic} includes a cross-sectional schematic and an optical microscope image of a 3$\times$3 pixel section of the array. The absorbers consist of \SI{0.8}{\um} thick gold and \SI{9}{\um} thick electroplated bismuth layers providing high X-ray absorption efficiency up to around \SI{30}{keV} without adding too much heat capacity. Representative measured parameters for this array are critical temperature $T_c \sim \SI{85.7}{mK}$, normal state resistance $R_N \sim \SI{13.7}{m\ohm}$, thermal conductance $G \sim \SI{380}{pW/K}$, and heat capacity $C \sim \SI{2}{pJ/K}$. Pixels are labelled with a number based on multiplexing resonator frequency order, not physical position, and with odd numbers only (i.e. pixels 3 and 5 are frequency neighbors, and the 33 resonators available map to pixel labels ranging from 1 to 65).

\subsection{Measurement}
The array was mounted in an ADR cryostat (bath temperature at \SI{65}{mK}) and biased with all pixels in series, within a range of 10\% to 15\% of each pixel's $R_N$. A total of 13 pixels showed satisfactory response to hard X-ray photons ($E > \SI{6}{keV}$). Known defects in the readout chain of this particular measurement setup account for the majority of the missing pixels.

In this study, the photons emitted by an X-ray tube source (maximum energy \SI{25}{keV}) and then scattered from a metal foil were collected. Previous experiments~\cite{Guruswamy2020} have demonstrated our hard X-ray TESs with a similar pixel design have a linear response up to \SI{30}{keV}, and a typical energy resolution of \SIrange{20}{40}{eV} at these photon energies and count rates up to 5 counts per pixel per second.
Previous measurements were done with an X-ray collimator in place to reduce off-absorber photon incidence; in this study we chose to continue without a collimator due to the low total photon incidence rate expected.

Three datasets were acquired, one of fluorescence from copper (Cu $K\alpha \sim \SI{8.0}{keV}$) and two of ruthenium (Ru $K\alpha \sim \SI{19.2}{keV}$). A \SI{300}{\um} thick silicon wafer was interposed between the foil and the detectors, and the X-ray tube cathode current was adjusted to maintain the detected photon flux at approximately 10 counts per second total across all working pixels of the entire array. The TES response is continuously sampled but only saved based on a triggering system -- when the signal from a pixel crosses a fixed threshold, a ``pulse record'' of fixed length is created and saved along with a microsecond-precision timestamp. With a record length of 2048 points sampled at \SI{62.5}{kHz} (total time \SI{32.7}{ms}), our choice to keep the photon incidence rate to below 10 counts per second for the whole array made it very likely that only one pixel is triggered at any one time, with minimal photon pile-up. This increases our chance of being able to isolate the response of all pixels to a single photon event without contamination from other coincident events.

To study crosstalk regardless of its magnitude, a photon event in any single pixel was set to trigger simultaneous acquisition by all pixels, known as ``group triggering" mode. This way if the triggering pixel (perpetrator) generated any spurious signal in another pixel (victim), this would be recorded even if below the typical threshold for acquisition in the victim. Acquisition continued until approximately \num{e6} group-triggered events were collected in total ($\sim \num{e4}$ real photon events per pixel, or several hours of data collection).

\subsection{Analysis}

\begin{table}[htb]
\caption{Selection criteria for perpetrator photon events to be used in crosstalk analysis\label{tab1}}
\begin{tabular}{lll}
\toprule
Parameter name & symbol & Filter criterion \\
\midrule
Peak value & $\max(x)$ & $> 0.05$ \\
Peak index & $\argmax(x)$ & $> 0$, $< 100$ \\
Pre-trigger standard deviation & $\sigma_{x,\text{pretrigger}}$ & $\le$ 5 $\mad(\sigma_{x,\text{pretrigger}}$) \\
Post-peak gradient & $\tod{x}{t}_\text{posttrigger}$ & $\le 5 \mad\left(\tod{x}{t}_\text{posttrigger}\right) $ \\
\bottomrule
\end{tabular}
\end{table}

All pulse processing and analysis was done offline. To determine the response in each victim pixel to a photon event in the perpetrator, we begin by selecting only those pulse records in the perpetrator that contain ``good" events, i.e. records that did not contain photon pile-up, tails from previous events, or spikes in noise that triggered the acquisition. \Cref{tab1} summarizes the selection criteria. Events that fail any of the selection rules are considered ``bad" and discarded before proceeding to subsequent analysis steps. Due to the low photon rate used, this was typically less than 10\% of saved pulse records. The ``median absolute deviation" $\text{mad}(x)$ is used instead of the sample standard deviation in some cases and is defined as
\begin{equation}
    \mad(\vec{x}) = \median \abs{\vec{x} - \median(\vec{x})}
\end{equation}
This combines the median, an estimator of mean insensitive to outliers, with a robust estimate of deviation, the absolute value, and ensures that random rare high energy photons or noise glitches do not overly influence the results. The pre-trigger standard deviation and post-trigger gradient are calculated based only on the portion of the saved record before and after the pulse peak, respectively. In our system all TES data are stored and analyzed in units of our demultiplexed SQUID amplifier response, as multiples of the flux quantum $\Phi_0$.

The ``good" events retrieved from the perpetrator pixel were manually inspected and seen to be mostly composed of the $K\alpha$ and $K\beta$ emissions from the foils. With perpetrator events selected, we next consider all traces from the victim pixels with timestamps matching the selected ``good" traces in the perpetrator pixel -- these correspond to potential crosstalk events. Any of these traces that contain a direct response from these victim pixels to a photon are filtered out by ensuring the peak value $\max(x) < 0.05$.
Finally, the crosstalk response is estimated by averaging the selected traces. This procedure was repeated for every perpetrator and victim combination in each dataset.

\section{Results and Discussion}\label{results}

\begin{figure}[htb]
\centering
\includegraphics[width=119mm]{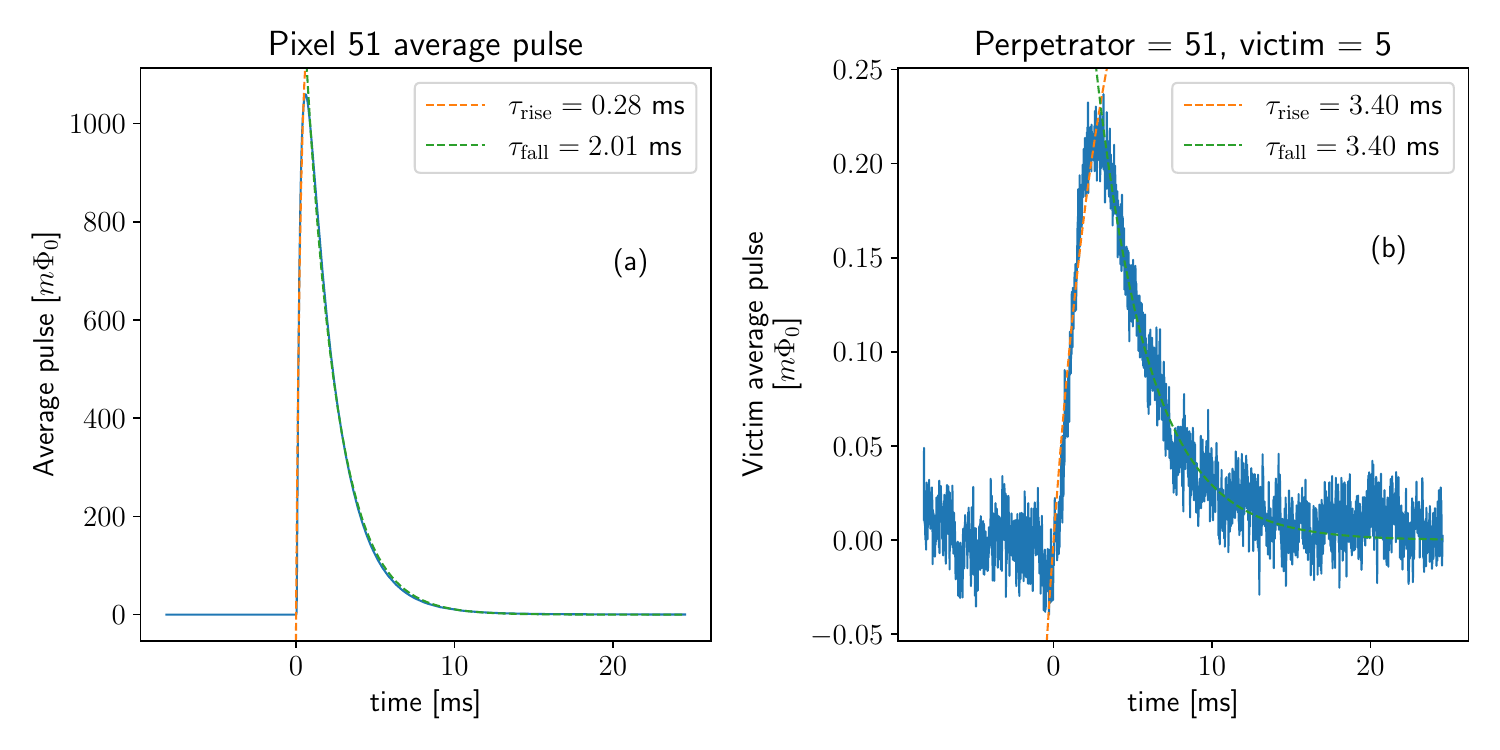}
\caption{(a) Average pulse for direct photon events in pixel 51 and (b) Pixel 5 average response to perpetrator pixel 51 photon events. Both measured with a Ru foil photon source. In both plots the dashed lines and legend text indicates the fitted rise and fall times for a single-exponential pulse model. These times are different for the original pulse but are nearly equal for the crosstalk response.\label{decay_time}}
\end{figure}

\Cref{decay_time}(a) shows the average pulse recorded for pixel 51, labelled with the fitted rise and fall times from a simple single-exponential pulse model of the form $x_\text{rise}(t) = A_\text{rise}\left[1 - \exp(-t/\tau_\text{rise})\right]$, $x_\text{fall}(t) = A_\text{fall}\left[\exp(-t/\tau_\text{fall})\right]$.
We find our pixels have a pulse rise time $\sim$ \SIrange{0.2}{0.3}{ms}, determined largely by the Nyquist inductance present in the TES bias circuit, and a fall time of $\sim$ \SIrange{1.5}{2}{ms}, which scales with the thermal conductance and heat capacity~\cite{Irwin2005}. \Cref{decay_time}(b) shows the corresponding average victim response in pixel 5 when pixel 51 has a photon event. A clear pulse-like response obeying causality (never occurring before the triggering event) is visible, and fitting this response with the same pulse model demonstrates that unlike the original event in pixel 51, the crosstalk response in pixel 5 has $\tau_\text{rise} \sim \tau_\text{fall}$, with both timescales similar to the thermally-determined time of the original perpetrator pulse. We suggest this crosstalk response is therefore mediated by a thermal mechanism, where some of the photon energy absorbed in a pixel is travelling through the frame of the TES array to cause a small response in neighbour pixels. The magnitude of the effect is small (less than 0.1\%) but still comparable to energy resolving power of the array ($E/\Delta{}E > 1000$) and so could plausibly be causing a negative effect on energy resolution. The effect should be proportional to the photon incidence rates, and so may explain a portion of the energy resolution degradation seen at high count rates from arrays of this design~\cite{Guruswamy2020}.

\begin{figure}[htb]%
\centering
\includegraphics[width=119mm]{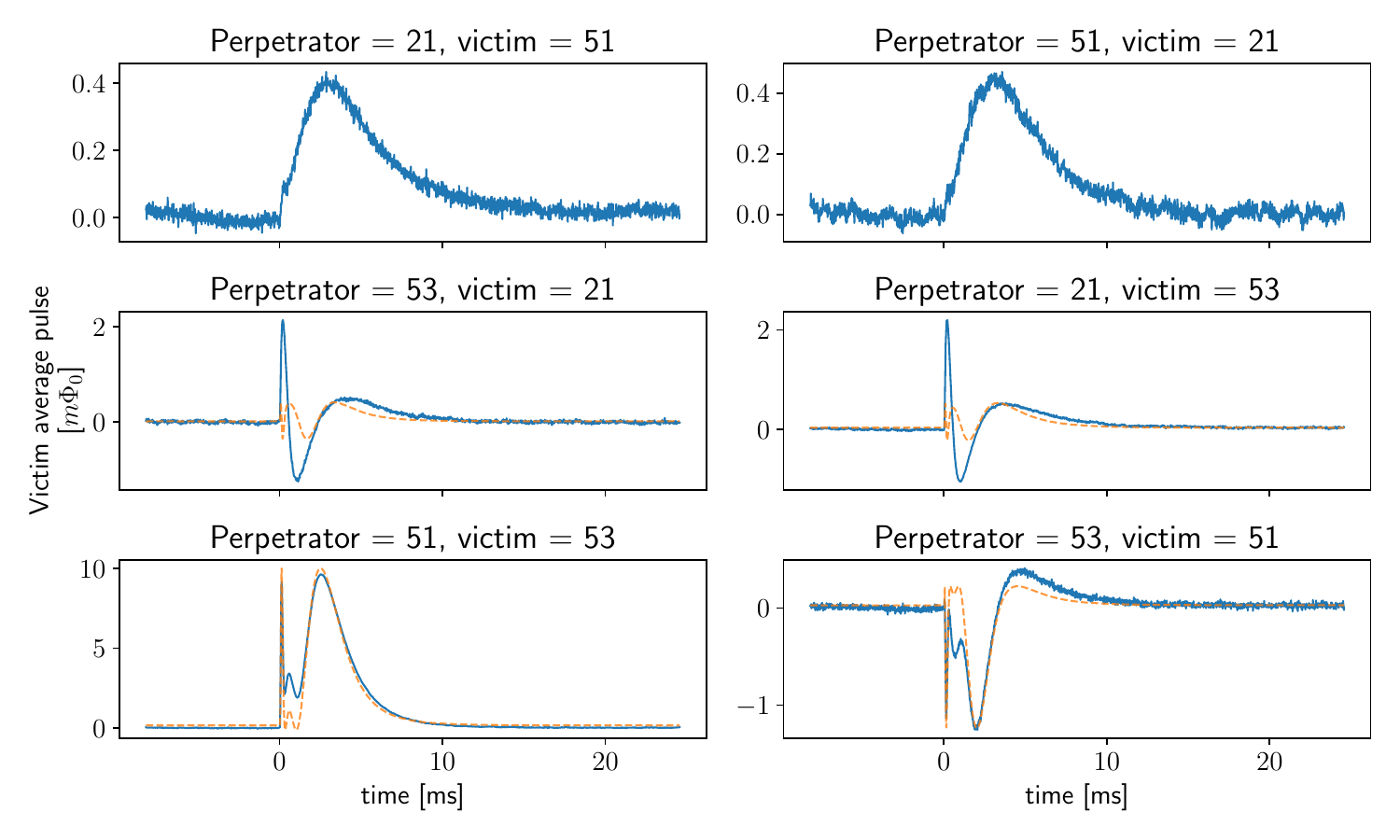}
\caption{Selected average victim pulse records (solid lines) with various perpetrator pixels. All measured with a Ru foil photon source. Pixel pair (21, 51) illustrate symmetric (on exchange of victim and perpetrator) thermal crosstalk, pair (21, 53) illustrate symmetric electrical crosstalk, and pair (51, 53) illustrate asymmetric electrical crosstalk. The middle and bottom row also include (dashed lines) a calculation of the expected crosstalk shape based on a sinusoidal modulation of the perpetrator pulse shape and fitted amplitude, due to our flux-ramped SQUID readout.\label{victim_examples}}
\end{figure}

Additional averaged pulse responses for various perpetrator-victim combinations are shown in \cref{victim_examples}. Some show a small pulse-like response as in \cref{decay_time}, however others show a faster oscillatory response, in some cases of significant magnitude. This kind of response is similar to the expected electrical crosstalk in frequency domain multiplexing readouts~\cite{Mates2019,Vaccaro2022}. Unlike a thermal response which is always positive and generally quite symmetric, the electrical response can sometimes be seen to invert in polarity when exchanging perpetrator and victim, for example in the pixel pair (51, 53). We hypothesize that the oscillatory responses which are largely symmetric between perpetrator and victim (e.g. pixels 21 and 55) are due to a DC electrical coupling, either on the TES array or in the low-frequency readout wiring before the multiplexing chip; and the oscillatory responses which invert (e.g. pixels 51 and 53) are due to a microwave-frequency coupling which depends on the frequency ordering of the pixels. This hypothesis is also in agreement with the model for flux-ramp modulated SQUID readout microwave-frequency crosstalk~\cite{Mates2019}, specifically that the crosstalk pulse resembles a sinusoidal modulation of the perpetrator pulse. A plot of $\Phi = \chi \sin((2\pi/\Phi_0)(\Phi_\text{perp}-\Phi_\text{vict})) + c$, with $\chi$ and $c$ obtained by fitting, is overlaid on the middle and bottom row of Fig. 2, showing agreement with the crosstalk pulse shape for the pixel pair (51, 53) but not for the pixel pair (21, 53).

Overall we find most combinations of perpetrator and victim show a small thermal crosstalk signature at least 1000 times smaller than the perpetrator pulse response. However a few specific combinations of pixels show significant electrical crosstalk -- these exceptions are therefore probably associated with specific local defects in wiring or multiplexing channel.

\begin{figure}[htb]%
\centering
\includegraphics[width=119mm]{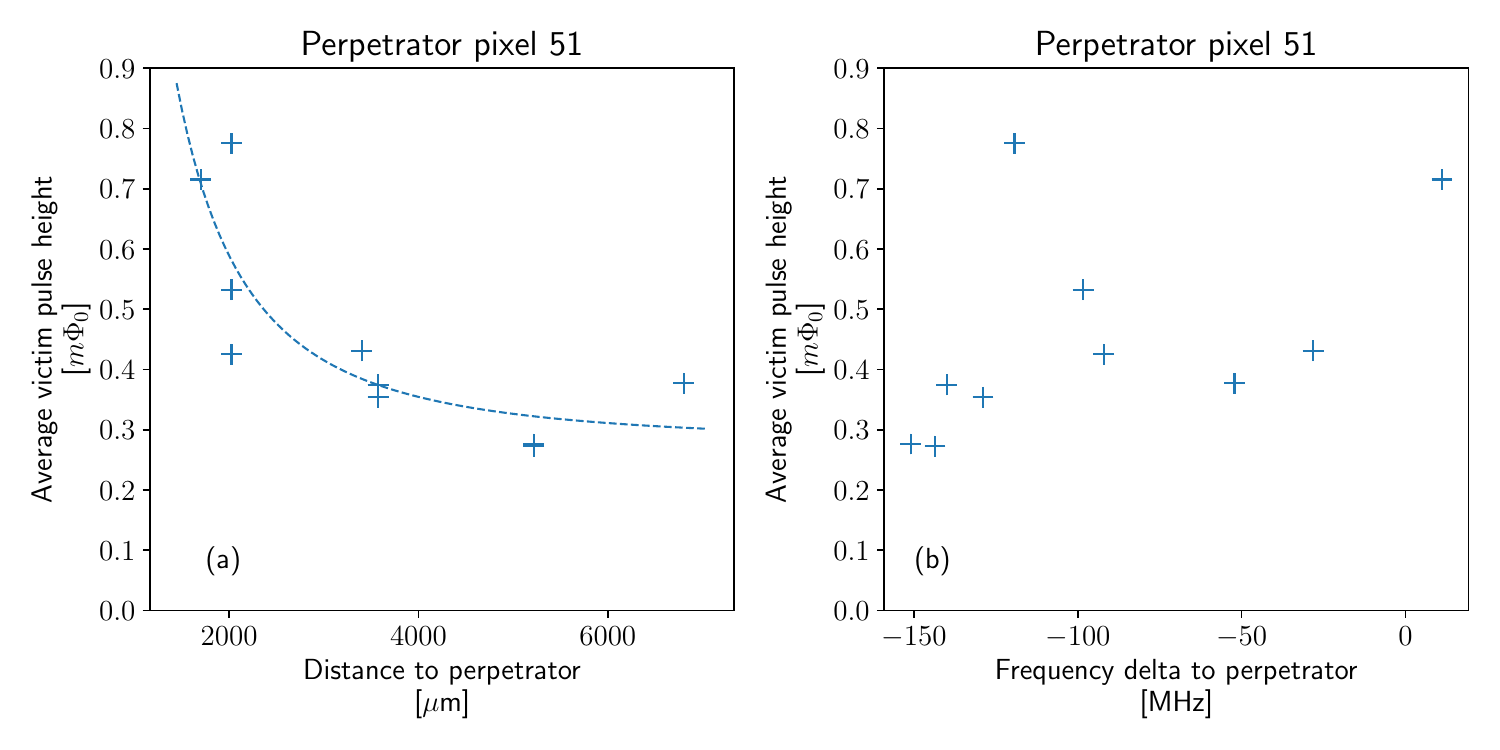}
\caption{(a) Average response magnitude as a function of physical distance between the perpetrator (pixel 51) and the victim pixels. The dashed line is $1/x^2$ scaled and shifted to match the data, illustrating the expected behavior for heat flow in 2D. The measured crosstalk is in qualitative agreement with this thermal model. (b) Average response magnitude as a function of the difference in frequency between the perpetrator and the victim readout resonators. No significant correlation is apparent.\label{distance_freq_correlation}}
\end{figure}

An exploration of crosstalk magnitude against physical pixel location indicates a weak correlation between the physical distance of the victim pixels to the perpetrator pixel and the average victim response, matching the observation of weak thermal crosstalk between most pixels in the array.
These correlations are explored in \cref{distance_freq_correlation}. The first panel shows that after excluding an outlier (electrical crosstalk with pixel 53), the victim pixel responses to photon events in pixel 51 are weakly correlated with physical distance. The functional dependence on distance appears consistent with a $1/x^2$ relationship, as would be expected for thermal conduction in a 2D film, and is similar to other experimental measurements of thermal crosstalk~\cite{Iyomoto2008a,Miniussi2020}. In contrast, there is no obvious correlation with of crosstalk magnitude with the distance in frequency-space of the microwave resonators corresponding to each pixel.
We found the results were consistent across all three datasets, even with a different average photon energy being absorbed (Cu vs Ru) and after a cryostat temperature cycle. This includes particular pixel combinations showing crosstalk with the same signature shape each time, supporting the repeatability and robustness of our method.

As well as applying these methods to understand crosstalk in the larger TES arrays in use in our scientific instruments, work to understand and address the physical origins of all crosstalk mechanisms identified in this array is ongoing. Beyond addressing defects, we anticipate thermal crosstalk may be addressed by improving the array mounting. Currently these chips are secured at the edges by copper clamps onto a device holder surface. To help thermalize the array, a gold (Au) ground plane is present on the top surface of the array, including in between pixels. This ground plane is then thermally anchored to the device holder surface via Au wire-bonds all around the perimeter of the chip. It has already been known that without these wire-bonds, the contact between the chip and mounting surface is sometimes insufficient to keep the devices at the measured cryostat temperature. It is possible that with further improvements to the mounting scheme and an increased thermal conductance to the cryostat bath, we can reduce the effect of photon events on the local temperature experienced by each pixel. Other reports~\cite{Miniussi2020} have shown the success of additional physical features on the TES chip such as muntins and extra metallic layers (front or back) in minimizing thermal crosstalk.

\section{Conclusion}\label{conclusion}
In this work we studied a 24-pixel hard X-ray TES microcalorimeter array fabricated at ANL with the objective of developing a procedure to investigate crosstalk effects without single-pixel illumination capabilities. We accomplished this by tuning the incident photon rate so that simultaneous events across the array were rare, recording all pixels when a threshold was triggered (``group triggering"), and then by averaging the synchronous response of the victim pixels when a selected perpetrator absorbed a photon. We found the signatures of thermal crosstalk (a slow pulse) and electrical crosstalk (a fast oscillatory response) in the shape of the victim responses. After averaging the magnitude of the crosstalk was measurable even at levels smaller than the typical readout noise of a pixel and showed a weak correlation with the physical distance on the chip between perpetrator and victim pixels. This, together with the shape of the response from the victim pixels and the observation of no significant relationship between the resonator frequency of each pixel and the average victim response, suggests this array design has overall a slight systematic issue with thermal crosstalk but not electrical crosstalk. However where electrical crosstalk is observed, in certain specific pixel combinations only and likely associated with defects, the magnitude can be quite significant.

Crosstalk can have a very significant negative impact on the performance of an instrument based on TESs, especially in the case of synchrotron applications, where large arrays operated at high photon count rates are crucial to effective science. In future studies, we will use the methods described here to measure and evaluate mitigations aimed at minimizing crosstalk to maintain optimal energy resolution performance throughout the entire photon energy range of operation in the larger 100-pixel arrays deployed at our beamline instrument.

\bmhead{Acknowledgments}
This research used resources of the Advanced Photon Source, a U.S. Department of Energy (DOE) Office of Science user facility, operated for the DOE Office of Science by Argonne National Laboratory under Contract No. DE-AC02-06CH11357. Work performed at the Center for Nanoscale Materials, a U.S. Department of Energy Office of Science User Facility, was supported by the U.S. DOE, Office of Basic Energy Sciences, under Contract No. DE-AC02-06CH11357.

\bibliography{aps_crosstalk_study_arxiv}

\end{document}